\documentclass[12pt,reqno]{article}
\usepackage{jheppub}
\usepackage{amsmath}
\usepackage{braket}
\usepackage{hyperref}
\usepackage{xfrac}
\usepackage{url}
\makeatletter
\newsavebox{\@brx}
\newcommand{\llangle}[1][]{\savebox{\@brx}{\(\m@th{#1\langle}\)}%
  \mathopen{\copy\@brx\kern-0.5\wd\@brx\usebox{\@brx}}}
\newcommand{\rrangle}[1][]{\savebox{\@brx}{\(\m@th{#1\rangle}\)}%
  \mathclose{\copy\@brx\kern-0.5\wd\@brx\usebox{\@brx}}}
\makeatother

\begin{document}

\preprint{SAGEX-20-16-E }
\title{Gravitational shock waves and scattering amplitudes}
\author{Andrea~Cristofoli}
\affiliation{Niels Bohr International Academy and Discovery Center\\ 
The Niels Bohr Institute, University of Copenhagen\\
Blegdamsvej 17, DK-2100 Copenhagen, Denmark}
\emailAdd{a.cristofoli@nbi.ku.dk}
\keywords{Scattering Amplitudes, General Relativity}
\date{\today}
\abstract{We study gravitational shock waves using scattering amplitude techniques. After first reviewing the derivation in General Relativity as an ultrarelativistic boost of a Schwarzschild solution, we provide an alternative derivation by exploiting a novel relation between scattering amplitudes and solutions to Einstein field equations. We prove that gravitational shock waves arise from the classical part of a three point function with two massless scalars and a graviton. The region where radiation is localized has a distributional profile and it is now recovered in a natural way, thus bypassing the introduction of singular coordinate transformations as used in General Relativity. The computation is easily generalized to arbitrary dimensions and we show how the exactness of the classical solution follows from the absence of classical contributions at higher loops. A classical double copy between gravitational and electromagnetic shock waves is also provided and for a spinning source, using the exponential form of three point amplitudes, we infer a remarkable relation between gravitational shock waves and spinning ones, also known as gyratons. Using this property, we infer a family of exact solutions describing gravitational shock waves with spin. We then compute the phase shift of a particle in a background of shock waves finding agreement with an earlier computation by Amati, Ciafaloni and Veneziano for particles in the high energy limit. Applied to a gyraton, it provides a result for the scattering angle to all orders in spin.}

\maketitle
\section{Introduction}
The study of General Relativity using scattering amplitudes techniques is in a golden era thanks to state of the art computations for interacting black holes and the possibility to relate classical gravitational observables to scattering amplitudes \cite{Damouruno,Poul,Cheung,Kosower,Bernuno,Berndue}. Nowadays, the literature is vast and includes different approaches to deal with post-Newtonian and post-Minkwoskian black holes \citep{Damourdue,Cristofoliuno,Paolodue,Cristofolidue,Paolouno,Bautista4,Solon,Martinez,Bjerrum,Kalinuno,Porto,
Blumlein1,Maier}, including also classical spin effects for Kerr black holes \cite{Vinesuno,Vinesdue,Guevarauno,Guevaradue,Nima,Maybeeuno,Vinestre,Damgaard,Aoude,Haddad,Nathan,Chung,Bernspin,Kalindue,Chung1,Chung2,Chung3} and tidal effects \cite{CheungSolon,Bini}. The existence of this literature might seem surprising, given that we are trading General Relativity for an even more complicated quantum gravitational system and its classical limit. However, the introduction of concepts such as unitarity and double-copy \cite{Doublecopy,Donal1,Donal2,Donal3,Donal4,Donal5,Donal6,Donal7,Donal8,Carrasco1,Carrasco2,Carrasco3,Henrik,Bautista3,Goldberger1,Goldberger2,Goldberger3,Goldberger4,Goldberger5,Plefka1,Plefka2,Plefka3,Adamo} has made possible not only the computation of observables relevant for LIGO/Virgo \cite{Buonanno} but also to unravel new structures in classical field theory, proving that quantum mechanics can help us in elucidating the essence of classical physics. In fact, the EOB approach \cite {EOB1}, which led to accurate models of gravitational wave signals for a binary system, was inspired by these ideas \cite{EOB2}. Along this line, this paper describes perturbative solutions in General Relativity using the scattering amplitude approach recently developed by Kosower, Maybee and O'Connell \cite{Kosower}. We focus on the Aichelburg-Sexl metric describing a gravitational shock wave sourced by a massless particle \cite{Aichelburg}. Derived almost simultaneously by Aichelburg, Sexl, Penrose \cite{Penrose} and Bonnor \cite{Bonnor}, it has been central to our understanding of graviton dominance in high energy scattering \cite{tHooft}, and in the past years it has been studied in different settings \cite{Dray, Kabat, Ortaggio1, Ortaggio2, Ortaggio3}. As we will see, an alternative derivation is also possible using a novel relation between perturbative solutions to Einstein's field equations and scattering amplitudes. Several authors have conjectured a similar connection and in the case of a static massive source, it has led to the computation at second order in $G_{N}$ of the Schwarzschild\footnote{See also \cite{Goldberger} for a derivation up to $G^{3}_{N}$ using EFT methods.} and Kerr-Newman solution  \cite{Arnau, Duff, Ira, Holstein}. However, the lack of a covariant framework has made it impossible to treat more general cases such as those described by an energy momentum tensor sourced by massless particles with spin. This work is a step toward this direction. We start by first reviewing the original derivation by Aichelburg and Sexl of a gravitational shock wave employing an ultrarelativistic boost of a Schwarzschild solution. We then present a relation between classical solutions in General Relativity and the classical part of three point functions from quantum field theory. Using a massless particle coupled to a graviton, we derive the complete Aichelburg-Sexl metric as an exact solution to Einstein field equations. The region where the radiation is localized has a well known distributional profile and it is now recovered from the amplitude itself, bypassing the introduction of singular coordinate transformations as used in General Relativity. We also generalize the computation to arbitrary $D$ dimensions finding agreement in the literature \cite{Vega, Ferrari} with the ultrarelativistic boost of the so called Tangherlini metric \cite{Emparan, Gustav}. We then extend the classical double copy for static black holes to gravitational shock waves, showing that their single copy is described by electromagnetic ones. For a spinning source, using the exponential form of three point amplitudes, we infer a remarkable relation between gravitational shock waves and spinning ones, also known as gyratons. From this, we obtain solutions describing spinning gravitational shock waves directly from the spinless case, avoiding the use of ultrarelativistic boosts on Kerr black holes \cite{Kerrboost}. To our knowledge, the existence of such a relation between exact solutions in General Relativity was previously unknown. Interestingly, this relation resembles the Newman-Janis algorithm \cite{Newman} which provides a Kerr solution from a complex deformation of  Schwarzschild, recently studied by Arkani-Hamed, Huang and O'Connell in \cite{Nima}. We then compute the phase shift of a particle in a background of shock waves, finding agreement with earlier computations for particles in the high energy limit \cite{ACV1,ACV2}. Applied to a gyraton, it provides a result for the scattering angle valid to all orders in the spin.\\
We will work throughout in natural units and in mostly negative signature.
\section{The Aichelburg-Sexl metric}
Aichelburg and Sexl derived for the first time an exact solution to Einstein field equations describing the gravitational field generated by a massless particle \cite{Aichelburg}. Their procedure employed the use of an ultrarelativistic boost of a Schwarzschild solution, previously used by D'Eath to address the scattering of two ultrarelativistic black holes \cite{DEath}. Let us review their original derivation. We start by introducing the Schwarzschild metric in isotropic coordinates \cite{Weinberg}
\begin{equation}
\label{isotropic}
d s^{2}=\frac{(1-A)^{2}}{(1+A)^{2}} \mathrm{d} t^{2}-(1+A)^{4}\left(\mathrm{d} x^{2}+\mathrm{d} y^{2}+\mathrm{d} z^{2}\right) \quad , \quad A=\frac{mG_{N}}{2 \sqrt{x^2+y^2+z^2}}
\end{equation}
and we decompose it as
\begin{equation}
d s^{2}=dt^2-dx^2-dy^2-dz^2+\bigg[\frac{(1-A)^{2}}{(1+A)^{2}}-1 \bigg]\mathrm{d} t^{2}-\bigg[(1+A)^{4}-1\bigg]\left(\mathrm{d} x^{2}+\mathrm{d} y^{2}+\mathrm{d} z^{2}\right)
\end{equation}
If we apply a Lorentz transformation to eq.(\ref{isotropic}) on the $x$ -direction,
\begin{equation}
t=\frac{\bar{t}-v \bar{x}}{\sqrt{1-v^2}} \quad , \quad  
x=\frac{\bar{x}-v\bar{t}}{\sqrt{1-v^2}} \quad , \quad  y=\bar{y} \quad , \quad z=\bar{z}
\end{equation}
the previous line element changes to
\begin{multline}
\label{newline}
d s^{2}=d\bar{t}^2-d\bar{x}^2-d\bar{y}^2-d\bar{z}^2+\bigg[\frac{(1-A')^{2}}{(1+A')^{2}}-1 \bigg] \frac{(d\bar{t}-v d\bar{x})^2}{1-v^2}\\
-\bigg[(1+A')^{4}-1\bigg]\left(\frac{(d\bar{x}-vd\bar{t})^2}{1-v^2}+\mathrm{d} \bar{y}^{2}+\mathrm{d} \bar{z}^{2}\right)
\end{multline}
\begin{equation}
\label{newA}
A'=\frac{m G_{N} \sqrt{1-v^2}}{2\left\{(\bar{x}-v \bar{t})^{2}+\left(1-v^{2}\right)\left(\bar{y}^{2}+\bar{z}^{2}\right)\right\}^{1 / 2}}
\end{equation}
\newline
We write $m= p \sqrt{1-v^2}$ and expand (\ref{newline}-\ref{newA}) around $v=1$ for a fixed value of $p$ to find
\newline
\begin{equation}
ds^2=d\bar{t}^2-d\bar{x}^2-d\bar{y}^2-d\bar{z}^2-\frac{4pG_{N}}{|\bar{t}-\bar{x}|} (d\bar{t}-d\bar{x})^2 \quad , \quad \bar{x} \neq \bar{t}
\end{equation}
\newline
In order to include also the missing region given by $\bar{x}=\bar{t}$, Aichelburg and Sexl proposed a coordinate transformation which becomes singular in the limit for $v=1$
\newline
\begin{equation}
\label{sing}
\begin{array}{l}
x^{\prime}-v t^{\prime}=\bar{x}-v \bar{t} \\
x^{\prime}+v t^{\prime}=\bar{x}+v \bar{t}-4 p G_{N} \log [\sqrt{(\bar{x}-v \bar{t})^{2}+\left(1-v^{2}\right)}-(\bar{x}-\bar{t})]
\end{array}
\end{equation}
Using the following relation 
\begin{equation}
\lim_{v \rightarrow 1}\bigg[ \frac{1}{\sqrt{(x'-vt')^2+(1-v^2) \rho}}-\frac{1}{\sqrt{(x'-vt')^2+(1-v^2)}}\bigg]=-2 \delta(t'-x') \log(\rho)
\end{equation}
\newline
the line element assumes the usual form of an impulsive pp-wave
\begin{multline}
\label{nomod}
ds^2=dt'^2-dx'^2-dy'^2-dz'^2 +4pG_{N}\delta(t'-x')\log(y'^2+z'^2)  (dt'-dx')^2
\end{multline}
The latter defines a global solution given by two copies of Minkwoski space connected by a singularity along a light cone coordinate. Among the relevant properties of this solution we can notice that in going from eq.(\ref{isotropic}) to eq.(\ref{nomod}) we have changed the algebraic type of the Weyl tensor from Petrov type D to the radiative type N \cite{Petrov}, a property first discovered by Pirani \cite{Pirani}. Moreover, from the computation of the associated Einstein tensor we can infer that the energy momentum tensor is simply that of a massless particle, thus confirming the physical interpretation of the metric.

\section{Gravitational shock waves from scattering amplitudes}
The idea to perturbatively solve Einstein field equations using quantum field theory techniques dates back to a paper by Duff \cite{Duff} where the Schwarzschild solution was derived up to $G^{2}_{N}$ order. After, several authors used known relations among off-shell scattering amplitudes and form factors so as to include quantum effects in the latter, confirming the same results for the classical part \cite{Holstein}. Both approaches require the knowledge of the Einstein-Hilbert action expanded around a fixed background which becomes intractable already after few iterations in the coupling $k=\sqrt{32 \pi G_{N}}$. In order to have a better control on the complexity of the calculation, it would be desirable to relate scattering amplitudes directly to the metric tensor in the same way as these have been related to classical observables in \cite{Kosower}. To this end we start by considering a Riemannian manifold and a Minkwoskian background. We then introduce an off-shell continuation of the second quantized solution to the linearized Einstein field equations
\begin{equation}
\label{operator}
\hat{h}_{\mu \nu}(x) = \frac{k}{2} \sum_{\lambda} \int d\Phi_{off}(q)\bigg[  \epsilon^{\lambda}_{\mu \nu}(q)\hat{a}^{\lambda}_{q} e^{-i q \cdot x}+(\epsilon^{\lambda}_{\mu \nu})^{\dag}(q)(\hat{a}^{\lambda}_{q})^{\dag} e^{i q \cdot x} \bigg]
\end{equation}
To ensure the gauge dependence of the metric, the sum runs over longitudinal polarizations and the measure of integration used in \cite{Kosower} has been replaced with
\begin{equation}
\label{measures}
d\Phi_{on}(q)= \frac{d^Dq}{(2\pi)^D} 2 \pi \delta(q^2)\theta(q_{0}) \quad \rightarrow \quad d\Phi_{off}(q)= \frac{d^Dq}{(2\pi)^D} \frac{1}{q^2} 
\end{equation}
We have added the subscripts \textit{on-off} to denote that every integral carrying such measure of integration will lead to an integrand with a momentum which is respectively on-shell or off-shell. In our case, eq.(\ref{measures}) ensures the off-shellness of the graviton and the fact we are not looking for radiative modes of the metric tensor. The $i \epsilon$ prescription is implicitly assumed. We then propose the following wave-function describing our system in absence of interactions
\begin{equation}
\ket{\Psi_{\mathrm{in}}}= \int d\Phi_{on}(p)  \phi(p) \ket{p} \otimes \ket{0} \quad , \quad d\Phi_{on}(p)= \frac{d^Dp}{(2\pi)^D} \hat{\delta}^{(+)}(p^2-m^2)
\end{equation}
where $\hat{\delta}^{(+)}(p^2-m^2)=2 \pi \delta(p^2-m^2)\theta(p_{0})$. The state $\ket 0$ denotes the vacuum state of the gravitational field while $\phi(p)$ is a proper wave-packet describing the source. We now define the metric tensor satisfying the non linear Einstein field equations as 
\newline
\begin{equation}
\label{definitionmetric}
g_{\mu \nu}(x)=\eta_{\mu \nu}+h_{\mu \nu}(x)\quad , \quad h_{\mu \nu}(x) =   \bra{\Psi_{I}(t)}\hat{h}^{I}_{\mu \nu}(x)\ket{\Psi_{I}(t)}
\end{equation}
\newline
The operator $\hat{h}^{I}_{\mu \nu}(x)$ is defined as the action of $U_{int}(+\infty,t)$ on (\ref{operator}), while the state $\ket{\Psi_{I}(t)}$ is defined as the evolution at time $t$ of the initial state under $U_{int}(t,-\infty)$. We can now express (\ref{definitionmetric}) as follows 
\begin{align}
\label{equa}
h_{\mu \nu}(x) &=\bra{\Psi_{I}(t)}U_{\mathrm{int}}^{\dag}(+\infty,t)\hat{h}_{\mu \nu}(x) U_{\mathrm{int}}(+\infty,t)\ket{\Psi_{I}(t)} \\
&=\bra{_{\mathrm{in}}\Psi}U_{\mathrm{int}}^{\dag}(t,-\infty)U_{\mathrm{int}}^{\dag}(+\infty,t)\hat{h}_{\mu \nu}(x) U_{\mathrm{int}}(+\infty,t)U_{\mathrm{int}}(t,-\infty)\ket{\Psi_{\mathrm{in}}}\\
\label{equa3}
&=\bra{_{\mathrm{in}}\Psi} S^{\dag}\hat{h}_{\mu \nu}(x) S\ket{\Psi_{\mathrm{in}}}\
\end{align}
where we have the introduced the $S$ matrix of the system. In doing so, we have been able to relate the solution to the complete Einstein field equations with a plane wave operator, by encoding all non linearities in the $S$ matrix alone.
Using then $S=1+iT$ we can expand eq.(\ref{equa3}) neglecting for the moment terms proportional to $TT^{\dag}$
\begin{equation}
h_{\mu \nu}(x) =  i \bra{\Psi_{\mathrm{in}}} \left(\hat{h}_{\mu \nu}(x) T-T^{\dagger} \hat{h}_{\mu \nu}(x)\right)  \ket{\Psi_{\mathrm{in}}}
\end{equation}
From which
\begin{multline}
h_{\mu \nu}(x)=\frac{ik}{2} \sum_{\lambda} \int d\Phi_{off}(q) d\Phi_{on}(p) d\Phi_{on}(p') \phi(p) \phi^{\dag}(p') \times \\ [\bra{p' q^{\lambda}} T \ket{p} \epsilon^{\lambda}_{\mu \nu}(q) e^{- i q \cdot x}-\bra{p'}T^{\dag}\ket{p q^{\lambda}}(\epsilon^{\lambda}_{\mu \nu})^{\dag}(q) e^{iq \cdot x}]
\end{multline}
\begin{equation}
\label{almost}
=-k\sum_{\lambda} \int d\Phi_{off}(q) d\Phi_{on}(p) d\Phi_{on}(p') \: \mathrm{Im}\bigg[\phi(p) \phi^{\dag}(p')   \bra{p' q^{\lambda}} T \ket{p} \epsilon^{\lambda}_{\mu \nu}(q)e^{- i q \cdot x}\bigg]
\end{equation}
\newline
Matrix elements in eq.(\ref{almost}) usually describe on-shell scattering amplitudes thanks to the covariant measures which have a Dirac delta in each integrated momentum. Having assumed instead an off-shell integration measure for gravitons, the term in eq.(\ref{almost}) won't be an on-shell scattering amplitude but an off-shell three point function  given by
\begin{equation}
 \bra{p' q^{\lambda}} T \ket{p} = (2\pi)^D\delta^{D}(p-q-p')\: (\epsilon^{\lambda}_{\alpha \beta})^{\dag}(q) \mathcal{M}^{\alpha \beta}(p,p',q)
\end{equation}
We now choose harmonic coordinates which amounts to requiring the following identity to hold
\begin{equation}
\label{gauge}
\sum_{\lambda} \epsilon^{\lambda}_{\mu \nu}(q)(\epsilon^{\lambda}_{\alpha \beta})^{\dag}(q)= \frac{1}{2} (\eta_{\mu \alpha}\eta_{\nu \beta}+\eta_{\mu \beta}\eta_{\nu \alpha}-\frac{2}{D-2}\eta_{\mu \nu}\eta_{\alpha \beta})\equiv P_{\mu \nu \alpha \beta}
\end{equation} 
Using this, eq.(\ref{almost}) becomes
\begin{multline}
h_{\mu \nu}(x)=-k \int d\Phi_{off}(q) d\Phi_{on}(p) d\Phi_{on}(p') \: \mathrm{Im}\bigg[\phi(p) \phi^{\dag}(p') \times\\  (2\pi)^D\delta^{D}(p'+q-p) \:P_{\mu \nu \alpha \beta} \mathcal{M}^{\alpha \beta}(p,p',q) e^{-i q \cdot x} \bigg]
\end{multline}
We now proceed by making explicit the integration measure for the source particle. Integrating over $p'$ we obtain
\begin{multline}
h_{\mu \nu}(x)=-k \int d\Phi_{off}(q) \frac{d^D p}{(2\pi)^D} \frac{d^D p'}{(2\pi)^D} \hat{\delta}^{(+)}(p^2-m^2) \hat{\delta}^{(+)}(p'^2-m^2) \times \\ \mathrm{Im}\bigg[\phi(p) \phi^{\dag}(p')  (2\pi)^D\delta^{D}(p-q-p') \:P_{\mu \nu \alpha \beta} \mathcal{M}^{\alpha \beta}(p,p',q) e^{-i q \cdot x} \bigg]
\end{multline}
\begin{multline}
h_{\mu \nu}(x)=-k \int d\Phi_{off}(q) \frac{d^D p}{(2\pi)^D} \hat{\delta}^{(+)}(p^2-m^2)\hat{\delta}^{(+)}((p-q)^2-m^2) \times \\ \mathrm{Im}\bigg[\phi(p) \phi^{\dag}(p-q)   \:P_{\mu \nu \alpha \beta} \mathcal{M}^{\alpha \beta}(p,p',q)  e^{-i q \cdot x} \bigg]
\end{multline}
where it is implicitly assumed that  $\mathcal{M}^{\alpha \beta}(p,p',q)$ is constrained with $p'=p-q$. Making also explicit the off-shell integration measure we obtain,
\begin{multline}
h_{\mu \nu}(x)=-k \int \frac{d^D q}{(2 \pi)^D}\frac{1}{q^2} \int \frac{d^D p}{(2\pi)^D} \hat{\delta}^{(+)}(p^2-m^2) \hat{\delta}^{(+)}(q^2-2 q \cdot p) \times \\ \mathrm{Im}\bigg[\phi(p) \phi^{\dag}(p-q)   \:P_{\mu \nu \alpha \beta} \mathcal{M}^{\alpha \beta}(p,p',q) e^{-i q \cdot x} \bigg]
\end{multline}
For a wave-packet sharply peaked around a given momentum $p_{0}$\footnote{For further details, see \cite{Kosower}, Section 4.} we obtain
\begin{align}
\label{masterfor}
h_{\mu \nu}(x)=-k \int \frac{d^D q}{(2 \pi)^D}\frac{\hat{\delta}^{(+)}(q^2-2 q \cdot p_0) }{q^2}  \:\mathrm{Im}[P_{\mu \nu \alpha \beta} \mathcal{M}^{\alpha \beta}(p_0,p'=p_0-q,q) e^{-i q \cdot x}]
\end{align}
\newline
We are thus left with a remarkable relation between the classical metric tensor satisfying Einstein field equation and three point functions with an external graviton, valid both for massive and massless sources
\newline
\begin{equation}
\label{masterfor}
h_{\mu \nu}(x)=- k \int \frac{d^D q}{(2 \pi)^D}\frac{\hat{\delta}^{(+)}(q^2-2 q \cdot p_0) }{q^2} \\ \: \mathrm{Im}[ P_{\mu \nu \alpha \beta} \mathcal{M}^{\alpha \beta}(p_0,p'=p_0-q,q) e^{-i q \cdot x}]
\end{equation}
\newline
Let us consider the massless case. Taking advantage of this covariant relation, we can proceed to explore which space-time corresponds to a three point function with an off-shell graviton and a massless source. Based on what has been discussed in Section 2, this should correspond to a gravitational shock wave. We start at tree level from the interaction of a graviton with a massless source
\newline
\begin{equation}
\label{feyn}
\mathcal{M}^{\mu \nu}(p_{1},p_{2})= \frac{ik}{2} \big(p^{\mu}_{1}p^{\nu}_{2}+p^{\mu}_{2}p^{\nu}_{1}- \eta^{\mu \nu} p_{1} \cdot p_{2} \big)
\end{equation}
\begin{equation}
\label{projections}
P_{\mu \nu \alpha \beta } \mathcal{M}^{\alpha \beta} (p,q)=\frac{ik}{2}\bigg[2p_{\mu}p_{\nu}-p_{\mu}q_{\nu}-p_{\nu}q_{\mu}+\eta_{\mu \nu} p\cdot q \bigg]
\end{equation}
\newline
where in the last equation we have expressed the whole contributions in terms of $p^{\mu}$ and $q^{\mu}$, being the former the incoming momenta. The whole metric tensor depends on only two functions
\newline
\begin{equation}
\label{thetasource}
h_{\mu \nu}(x)=-\frac{k^2}{2}\bigg[2p_{\mu}p_{\nu}\Theta(x)-p_{\mu}\Theta_{\nu}(x)-p_{\nu}\Theta_{\mu}(x)+\eta_{\mu \nu} p^{\alpha} \Theta_{\alpha}(x)  \bigg]
\end{equation}
\begin{align}
\label{thetauno}
\Theta(x)&=\int \frac{d^Dq}{(2\pi)^D} \hat{\delta}^{(+)}(q^2-2 q \cdot p) \frac{\cos(q\cdot x)}{q^2}
\\
\label{thetadue}
\Theta_{\mu}(x)&=\int \frac{d^Dq}{(2\pi)^D} \hat{\delta}^{(+)}(q^2-2 q \cdot p) \frac{q_{\mu}}{q^2}\: \cos(q \cdot x)
\end{align}
\vspace{2mm}
\newline
In the classical limit we implement the limit for small $q$ by considering the integration domain where $
p \gg q$. This amounts to disregarding the Heaviside theta in eq.(\ref{masterfor}) as well as the $q^2$ term in its Dirac delta
\newpage
\begin{equation}
\hat{\delta}^{(+)}(q^2-2 q \cdot p) \quad \rightarrow \quad  2 \pi \delta(2 q \cdot p)
\end{equation} 
\newline
In this limit $\Theta_{\mu}(x)$ is vanishing being the integrand an odd and real valued function and we are thus left with the computation of $\Theta(x)$. Using then the integral representation for a Dirac delta together with the Schwinger parametrization we obtain
\begin{equation}
\Theta(x)=- i  \int_{\mathbb{R}} ds  \int_{\mathbb{R}_{+}} dt  \int \frac{d^D q}{(2\pi)^D}e^{-i q \cdot (x-2ps)+i q^2t}
\end{equation}
\newline
The latter is a complex Gaussian integral and its computation gives
\begin{equation}
\Theta(x)=-\bigg(\frac{i}{4 \pi}\bigg)^{\frac{D}{2}} \int_{\mathbb{R}} ds  \int_{\mathbb{R}_{+}} dt \: \frac{e^{-i \frac{(x-2ps)^2}{4t}}}{t^{\frac{D}{2}}}
\end{equation}
\newline
Expanding the square in the exponential,
\newline
\begin{align}
\Theta(x)&=-\bigg(\frac{i}{4 \pi}\bigg)^{\frac{D}{2}}  \int_{\mathbb{R}} ds  \int_{\mathbb{R}_{+}} dt \: \frac{e^{-i \frac{(x-2ps)^2}{4t}}}{t^{\frac{D}{2}}}
\\
&=-2 \pi  \bigg(\frac{i}{4 \pi}\bigg)^{\frac{D}{2}}  \int_{\mathbb{R}_{+}} \frac{dt}{t^{\frac{D}{2}}}e^{-\frac{ix^2}{4t}}\delta \bigg(\frac{p \cdot x}{t} \bigg) 
\\
&=-2 \pi  \bigg(\frac{i}{4 \pi}\bigg)^{\frac{D}{2}}  \delta(p \cdot x) \int_{\mathbb{R}_{+}} \frac{dt}{t^{\frac{D}{2}}} e^{\frac{-i x^2}{4y}} |t|
\end{align}
\newline
Changing variables to $t=\frac{1}{u}$, we get
\newline
\begin{equation}\label{arbitraryD}
\Theta(x)=-2 \pi  \bigg(\frac{i}{4 \pi}\bigg)^{\frac{D}{2}} \delta(p \cdot x) \int_{\mathbb{R}_{+}} du \: \frac{e^{\frac{-i u x^2}{4}}}{u^{\frac{6-D}{2}}}
\end{equation}
\newline
At this point, we should carefully distinguish the computation in $D=4$ from other dimensions. One can realize it by computing separately the two cases and by a comparison afterwords. We start from the case with $D=4$, 
\begin{equation}
\label{log}
\Theta(x)=\frac{1}{8 \pi}\delta(p \cdot x) \int_{\mathbb{R}_{+}} \frac{du}{u} e^{-\frac{iu x^2}{4}}
\end{equation}
In order to compute this integral, we consider its partie finie ($\mathrm{Pf}$) to find
\newline
\begin{equation}
\begin{aligned}
\Theta(x) &=\frac{1}{8 \pi} \delta(p \cdot x) \operatorname{Pf} \lim _{z \rightarrow 0} \int_{z}^{+\infty} \frac{d u}{u} e^{-\frac{i u x^{2}}{4}} \\
&=\frac{1}{8 \pi} \delta(p \cdot x) \operatorname{Pf} \lim _{z \rightarrow 0} \int_{1}^{+\infty} \frac{d u}{u} e^{-\frac{i z u x^{2}}{4}} \\
&=\frac{1}{8 \pi} \delta(p \cdot x) \operatorname{Pf} \lim _{z \rightarrow 0} E_{1}\left(\frac{z i x^{2}}{4}\right)
\end{aligned}
\end{equation}
\newline
where we have introduced the exponential integral $E_{1}(x)$. Using the Puiseux series
\begin{equation}
E_{1}(z)=-\gamma-\log z-\sum_{k=1}^{\infty} \frac{(-z)^{k}}{k k !} \quad , \quad |arg (z)<\pi|
\end{equation}
The result is
\begin{equation}
\Theta(x)=-\frac{1}{8 \pi}\delta(p \cdot x) \log(|x^2|)  \quad , \quad D=4
\end{equation}
\newline
As for the case $D \neq 4$, we evaluate it by first rescaling eq.(\ref{arbitraryD}),
\begin{equation}
\label{remain}
\Theta(x)=-2 \pi  \bigg(\frac{i}{4 \pi}\bigg)^{\frac{D}{2}} \delta(p \cdot x) \bigg( \frac{x^2}{4}\bigg)^{\frac{4-D}{2}} \int_{\mathbb{R}_{+}} du\: e^{-i u} u^{\frac{D-6}{2}}  
\end{equation}
After a Wick rotation, the remaining integral defines a Gamma function, from which
\newline
\begin{equation}
\Theta(x)=\frac{ \pi^{\frac{2-D}{2}} }{4} \frac{\Gamma(\frac{D-2}{2})}{D-4} \frac{\delta(p \cdot x)}{(x^2)^{\frac{D-4}{2}}} \quad , \quad D>4
\end{equation}
\newline
We can thus summarize our results,\footnote{One could also infer the $D = 4$ case from the following regularization $\frac{\Gamma(\frac{D-4}{2})}{x^{D-4}} \rightarrow -2 \log(x)$. This amounts to remove a divergent quantity from the metric tensor with a gauge transformation.}
\newline
\begin{equation}
\Theta(x)=\begin{cases}
   -\frac{1}{8 \pi}\delta(p \cdot x) \log(|x^2|)  \quad , \quad  D=4\\
    \\
     \frac{ \pi^{\frac{2-D}{2}} }{4} \frac{\Gamma(\frac{D-2}{2})}{D-4} \frac{\delta(p \cdot x)}{(x^2)^{\frac{D-4}{2}}}             \quad \: \: \:, \quad \text{otherwise}
\end{cases}
\end{equation}
\newline
Using eq.(\ref{thetasource}) we can read the metric tensor related to
a three point function with a massless source and an off-shell graviton.\newpage The final result is
\begin{equation}
\label{final}
h_{\mu \nu}(x)=
\begin{cases}
    4G_{N} p_{\mu} p_{\nu}  \delta(p \cdot x) \log (|x^2|) \quad \quad \: , \quad D=4
    \\
    -8\pi^{\frac{4-D}{2}}G_{N} p_{\mu} p_{\nu} \frac{\Gamma \big(\frac{D-2}{2}\big)\delta(p \cdot x)}{(D-4)(x^2)^{\frac{D-4}{2}}}  \quad , \quad \text{otherwise}
\end{cases}
\end{equation}
\newline
We will shortly argue that contributions from higher loops produce only divergences which are removed from the cut terms proportional to $TT^{\dag}$ in (\ref{equa3}). This procedure provides an exact solution to Einstein field equations already at linear order in $G_{N}$. In $D=4$ the line element reads
\begin{equation}
ds^2= g_{\mu \nu} dx^{\mu} dx^{\nu}=dt^2-dx^2-dy^2-dz^2+4G_{N} \delta(p \cdot x) \log (|x^2|) p_{\mu}p_{\nu}dx^{\mu}dx^{\nu}
\end{equation}
For a massless particle moving along the $x$ direction we recover the Aichelburg-Sexl metric (\ref{nomod}) for a gravitational shock wave 
\begin{equation}
ds^2=dt^2-dx^2-dy^2-dz^2+4p G_{N}\delta(t-x)\log(y^2+z^2) (dt-dx)^2 \
\end{equation}
In $D$ dimensions, the metric is in agreement with earlier computations describing the ultrarelativistic boost of the Schwarzschild-Tangherlini metric in $D$ dimensions \cite{Ferrari}. As for the coordinates associated with this metric, we notice that eq.(\ref{final}) satisfies the harmonic gauge condition, equivalent at linear order in $G_{N}$ with the linear harmonic,
\begin{equation}
\eta^{\alpha \beta}\Gamma^{\mu}_{\alpha \beta}=0 \quad \rightarrow \quad \partial_{\alpha}h^{\mu \alpha}=\frac{1}{2}\partial^{\mu}h
\end{equation}
which can be easily seen to be satisfied thanks to eq.(\ref{final}) being traceless. In $D=4$  
\begin{equation}
\partial_{\alpha}h^{\mu \alpha}=4G_{N}p^\mu p^{\alpha}\partial_{\alpha} \delta(p \cdot x) \log (|x^2|)+8G_{N}p^{\mu}p^{\alpha}\delta(p \cdot x)\frac{x_{\mu}}{x^2}=0
\end{equation}
with the same result in higher dimensions. This is consistent with the harmonic gauge choice made in eq.(\ref{gauge}). 
The advantages of this computation with respect to the derivation from classical General Relativity are several. The exactness of the solution already at linear in $G_{N}$ can now be explained in light of the absence of classical contributions to higher loops in three point functions with two massless scalars and a graviton \cite{Poul}. This is in contrast with the computation for a massive three point function where the classical part from higher loop orders is non vanishing and needed in order to reproduce the expansion of Schwarzschild in $G_{N}$ \cite{Holstein}. Remarkably, the distributional profile emerges in a natural way from the amplitude itself, with no need to introduce singular coordinate transformations as those in eq.(\ref{sing}). As we will see, this property is more general: it is valid also for gravitational shock waves carrying a spin dependence.

\section{A classical double copy for gravitational shock waves}
In the previous section we have shown a relation between perturbative solutions to Einstein field equations and scattering amplitudes. The latter can be introduced also for a gauge theory as classical electromagnetism. We start by introducing the following operator for a gauge potential
\begin{equation}
\hat{A}^{\mu}(x)=\sum_{\lambda} \int d\Phi_{off}(q)\bigg[  \epsilon^{\lambda}_{\mu}(q)\hat{a}^{\lambda}_{q} e^{-i q \cdot x}+(\epsilon^{\lambda}_{\mu})^{\dag}(q)(\hat{a}^{\lambda}_{q})^{\dag} e^{i q \cdot x} \bigg]
\end{equation}
Following the same steps seen before and working in Feynman gauge we can easily derive a relation between a gauge potential $A^{\mu}(x)$ and three point functions in scalar QED with an external photon,
\begin{equation}
A^{\mu}(x)=\int \frac{d^D q}{(2 \pi)^D}\frac{\hat{\delta}^{(+)}(q^2-2 q \cdot p_0) }{q^2} \\ \mathrm{Im}[\mathcal{M}^{\mu}(p_0,p'=p_0-q,q) e^{-i q \cdot x}]
\end{equation}
\newline
We now consider electromagnetic shock waves \cite{Lechner}. We find natural to relate these to a three point amplitude of a massless scalar particle coupled to a photon,
\begin{equation}
\mathcal{M}^{\mu}= -i e (2 p^{\mu}-q^{\mu})
\end{equation}
Using this, we can express the gauge potential $A^{\mu}(x)$ in terms of (\ref{thetauno}, \ref{thetadue}),
\begin{equation}
A^{\mu}(x)= -e\:  \bigg[2p^{\mu}\Theta(x)-\Theta^{\mu}(x)\bigg]
\end{equation}
The final result for an electromagnetic shock wave is
\begin{equation}
A^{\mu}(x)=
\begin{cases}
     \frac{e}{4\pi}\: 
     p^{\mu}   \delta(p \cdot x) \log (|x^2|)\: \: \: \: \: \: \: \quad , \quad D=4\\
    \\
     -\frac{e}{2\pi }\:  p^{\mu} \: \pi^\frac{4-D}{2} \frac{\Gamma \big(\frac{D-2}{2}\big)\delta(p \cdot x)}{(D-4)(x^2)^{\frac{D-4}{2}}}  \quad , \quad \text{otherwise}
\end{cases}
\end{equation}
\newline
We may now consider the classical double copy procedure shown in \cite{Donal3} in order to construct a solution in General Relativity. This amounts to the following replacement
\begin{equation}
e \rightarrow 16 \pi G_{N} \quad , \quad p^{\mu} \rightarrow p^{\mu} p^{\nu}
\end{equation}
Remarkably, this gives the correct gravitational shock wave in arbitrary $D$ dimensions of eq.(\ref{final}) showing that gravitational and electromagnetic shock waves are related by a classical double copy. This fact was shown to hold for static black holes \cite{Donal4, Donal6} and then generalized to  accelerating ones in \cite{Donal3}. Here, we have proven that a classical double copy is satisfied also by shock waves in gravity and gauge theories. 

\section{Spinning gravitational shock waves}
Having studied in depth the relation between massless particles and gravitational shock waves, we find natural to investigate the same relation for the case of a spinning source. As we will see, this leads to a family of classical solutions also known in the literature as gyratons \cite{Bonnorspin, Frolovspin}. For ease of discussion we restrict ourselves to the case $D=4$. In order to perform the computation, we take advantage of the exponential representation of three point functions for a spinning massive particle emitting a graviton \cite{Guevarauno, Aoude}
\newline
\begin{equation}
\label{Threeexp}
\mathcal{M}_{\mu \nu}^{S}= \mathcal{M}_{\mu \nu}\: e^{a \cdot q} \quad , \quad a^{\mu}= \frac{s^{\mu}}{m}
\end{equation} 
\newline
being $\mathcal{M}_{\mu \nu}$ the associated spinless three point amplitude and $s^{\mu}$ the spin vector of the source. In the massless limit we treat $a^{\mu}$ as a constant and express the spin vector using the representation of the Pauli-Lubanski pseudovector for a massless particle \cite{Maggiorebook}
\begin{equation}
\label{spin}
s^{\mu}=\frac{1}{E}W^{\mu} \quad , \quad W^{\mu}= \frac{1}{2} \epsilon_{\mu \nu \alpha \beta}S^{\nu \alpha}p^{\beta}
\end{equation}
Using this, the metric tensor reads
\newline
\begin{equation}
h^{S}_{\mu \nu}(x)= -\frac{k^2}{2}\bigg(2p_{\mu}p_{\nu}\Theta^{S}(x)-p_{\mu}\Theta^{S}_{\nu}(x)-p_{\nu}\Theta^{S}_{\mu}(x)+ \eta_{\mu \nu} \Theta^{S}_{\alpha}(x) p^{\alpha} \bigg)
\end{equation}
\begin{align}
\label{thetaspinuno}
\Theta^{S}(x)&=\int \frac{d^4q}{(2\pi)^4} \hat{\delta}^{(+)}(q^2-2 q \cdot p) \frac{\cos(q\cdot x)e^{ a \cdot q}}{q^2}
\\
\label{thetaspindue}
\Theta_{\mu}^{S}(x)&=\int \frac{d^4q}{(2\pi)^4} \hat{\delta}^{(+)}(q^2-2 q \cdot p) \frac{q_{\mu}}{q^2}\: \cos(q \cdot x) e^{ a \cdot q}
\end{align}
\newline
We can now prove a relation between spinless gravitational shock waves and gyratons. We restrict to the classical limit by considering the integration region where $
p \gg q
$, thus
\newline
\begin{align}
\Theta^{S}(x)&=\frac{\Theta(x-ia)+\Theta(x+ia)}{2} \quad , \quad \Theta^{S}_{\mu}(x)&=\frac{\Theta_{\mu}(x-ia)+\Theta_{\mu}(x+ia)}{2}
\end{align}
\newline
We now consider the behavior of $\Theta(x)$ under a complex shift. Introducing $a=\sqrt{-a^{\mu}a_{\mu}}$ we obtain the following expression
\begin{equation}
\label{spin}
\Theta(x-ia)=  - \frac{1}{8 \pi}\delta(p \cdot x) \log(|x^2+a^2|)
\end{equation}
Eq.(\ref{spin}) is real valued due to the absence of linear terms in $a \cdot x$. This is ensured by the Dirac delta in $p \cdot x$ and the fact that the Pauli-Lubanski pseudovector and $a^{\mu}$ are proportional to $p^{\mu}$ \cite{Maggiorebook}. Thus,
\begin{equation}
\begin{cases}
\Theta^{S}(x)=\Theta(x-ia) \\
\Theta^{S}_{\mu}(x)=\Theta_{\mu}(x-ia)
\end{cases}
\quad \quad \rightarrow \quad  \quad \label{shiftproperty} h_{\mu \nu}^{S}(x)=h_{\mu \nu}(x-ia)
\end{equation}
\newline
Remarkably, thanks to the exponential form of the three point amplitude we can now read the metric tensor sourced by a massless spinning source directly from the spinless case using the shift $x^{\mu} \rightarrow x^{\mu}-ia^{\mu}$. To our knowledge, this property between spinless shock waves and gyratons was previously unknown and it relates to the exponential form of the energy momentum tensor for linearized Kerr black holes \cite{Vinesuno}, which is preserved after an ultrarelativistic boost. The line element is
\begin{equation}
ds^2=dt^2-dx^2-dy^2-dz^2+4G_{N} \delta(p \cdot x) \log(|(x^2+a^2|) p_{\mu}p_{\nu}dx^{\mu} dx^{\nu}
\end{equation}
In particular, for a spinning particle moving along the $x$ direction
\begin{equation}
ds^2=dt^2-dx^2-dy^2-dz^2+4G_{N}p \delta(x-t) \log (|y^2+z^2-a^2|) (dt-dx)^2
\end{equation}
in agreement with an earlier computation by Ferrari and Pendenza \cite{Kerrboost} describing the ultrarelativistic boost of a Kerr black hole. The derivation by a simple shift in $a^{\mu}$ is remarkable, since the same in classical General Relativity is much more complicated.\\ Interestingly, this procedure resembles the Newman-Janis algorithm \cite{Newman} which provides a Kerr solution from a complex deformation of Schwarzschild, this last recently studied by Arkani-Hamed, Huang and O'Connell in \cite{Nima}. As for the singularity at $y^2+z^2=a^2$, we interpret this as the remnant of the singularity in the equatorial plane.

\section{The scattering angle in the high energy limit}
The computation of geodesics in a gravitational shock wave background has been explored by several authors \cite{Baueruno,Bauerdue}. Since the whole space-time is Minkwoskian up to a region defined by a null light cone coordinate, geodesics are fully determined from the net change in momentum of a particle 
\begin{equation}
\Delta p_{0}^{\mu}=\frac{1}{2} \int_{\mathbb{R}} d \sigma \partial^{\mu} h_{\alpha \beta}(x(\sigma)) p_{0}^{\alpha} p_{0}^{\beta}
\end{equation}
where the subscript $0$ denotes the particle, while $\sigma$ the affine parameter of its world-line. To leading order in $G_{N}$ we assume free motion
\begin{equation}
x_{0}^{\mu}(\sigma)=p_{0}^{\mu} \sigma+b^{\mu} \quad, \quad b \cdot p_{0}=0 \quad, \quad b \cdot p=0
\end{equation}
being $b^{\mu}$ a covariant impact parameter and $p^{\mu}$ the momentum associated with the shock wave. The resulting change of impulse in $D=4$ reads
\newline
\begin{align}
\Delta p_{0}^{\mu}&=\frac{1}{2} p_{0}^{\alpha} p_{0}^{\beta} \int_{\mathbb{R}} d \sigma 8 G_{N} p_{\alpha} p_{\beta} \left[\delta\left(p \cdot x_{0}(\sigma)\right) \frac{x_{0}^{\mu}(\sigma)}{ x_{0}^{2}(\sigma)}\right]=\frac{4 G_{N} p\cdot p_{0}}{b \cdot b} \: b^{\mu}
\end{align}
\newline
Having computed the change of momentum experienced by the particle, we can compute the associated phase shift using
\begin{equation}
\sin (\theta)=\frac{\Delta p^{\mu}_{0} b_{\mu}}{p_{0}\: b} 
\end{equation}
where we have introduced $b=\sqrt{-b_{\mu}b^{\mu}}$. The result is
\begin{equation}
\sin (\theta)=\frac{4 G_{N} p \cdot p_{0}}{p_{0}\: b}
\end{equation}
Let's now consider the massless limit,
\begin{equation}
p \cdot p_{0}=p \:p_{0}-\vec{p} \cdot \vec{p}_{0}=2 p_{0}^{2} \quad , \quad  s=4 p_{0}^{2}
\end{equation}
If we now apply the small angle approximation we obtain
\begin{equation}
\theta=\frac{4 G_{N} \sqrt{s}}{b}
\end{equation}
in agreement with an earlier computation by Dray and t'Hooft \cite{Dray}. Interestingly, as shown by Amati, Ciafaloni and Veneziano, the same result agrees with the leading order scattering angle between particles in the high energy limit \cite{ACV1}. We can generalize this result including effects to all orders in spin using as a source the metric tensor for a gyraton derived in eq.(\ref{shiftproperty}). This provides the following result
\begin{equation}
\Delta p^{\mu}= \frac{4 G_{N} p \cdot p_{0} \: b^{\mu}}{a^2-b^2}
\end{equation}
The scattering angle in the massless limit and including effects to all order in spin reads
\newline
\begin{equation}
\theta=\frac{2 G_{N} \sqrt{s}}{b-a}+\frac{2 G_{N} \sqrt{s}}{b+a}
\end{equation}
\newline
Interestingly, this scattering angle in the high energy limit resembles a striking similarity with the all order in spin result by Vines \cite{Vinesuno} including the pole at $b=a$.
\vspace{2mm}
\newline
\section{Conclusion}
We have derived a relation between perturbative solutions to Einstein field equations and off-shell scattering amplitudes thanks to a covariant framework developed by Kosower, Maybee and O'Connell \cite{Kosower}. We have studied to which gravitational field corresponds a scattering amplitude with an off-shell graviton and two massless particles finding that the latter describes a gravitational shock wave also known as Aichelburg-Sexl metric \cite{Aichelburg}. The result has been easily generalized to arbitrary $D$ dimensions finding agreement with previous computation of $D$ dimensional shock waves in General Relativity \cite{Ferrari}. The advantage of this computation are several. We have been able to avoid singular coordinate transformations which were used in General Relativity to deal with the singular behavior of the gravitational field along a light cone coordinate. Remarkably, the distributional profile emerges in a natural way from the amplitude itself, while the exactness of the classical solution at linear in $G_{N}$ can now be explained in light of the absence of classical contributions at higher loops for three point functions with massless particles. We have also shown that a classical double copy is satisfied between gravitational and electromagnetic shock waves and for a spinning source, using the exponential form of three point amplitudes, we have inferred a remarkable relation between gravitational shock waves and spinning ones, also known as gyratons. Using this property, we have been able to infer solutions describing spinning gravitational shock waves directly from the spinless case, thus bypassing the derivation in General Relativity involving an ultrarelativistic boost of a Kerr black hole. We have computed the phase shift of a particle in a background of shock waves finding agreement with earlier computations for the scattering angle of particles in the high energy limit \cite{ACV1,ACV2}. Applied to a gyraton, it has provided a result to all orders in the spin.

\section{Acknowledgments}
I am sincerely thankful to Poul Damgaard, for his constant support and several useful conversations. I also thank Donal O'Connell, Paolo Di Vecchia and Marcello Ortaggio for useful conversations and comments on the manuscript. This work has been based partly on funding from the European Union’s Horizon 2020 research and innovation programme under the Marie Skłodowska-Curie grant agreement No. 764850 (“SAGEX”).
\vspace{2mm}
\newline

\end{document}